\documentclass[
twocolumn,
superscriptaddress,
amsmath,amssymb,
aps,
prl,
longbibliogtraphy
]{revtex4-1}

\DeclareUnicodeCharacter{25CF}{$\bullet$}

\usepackage{graphicx}
\usepackage[dvipsnames]{xcolor}
\usepackage{amssymb}
\usepackage{amsmath}
\usepackage{color}
\usepackage{bm}

\usepackage{xr}
\newcommand{\koff}{$k_{\textrm{off}}$~}

\begin{document}

\title{Topoisomerase DNA Binding Kinetics Differentially Regulates Decatenation and Torsional Simplification Rates}
\title{Role of DNA Binding Kinetics in Regulating Topoisomerase Simplification Activity}
\title{Binding Kinetics Oppositely Regulates type II Topoisomerase Relaxation and Decatenation Activities}

\author{Cleis Battaglia}
\affiliation{School of Physics and Astronomy, University of Edinburgh, Peter Guthrie Tait Road, Edinburgh, EH9 3FD, UK}
\author{Filippo Conforto}
\affiliation{School of Physics and Astronomy, University of Edinburgh, Peter Guthrie Tait Road, Edinburgh, EH9 3FD, UK}
\author{Yair Augusto Guti\'{e}rrez Fosado}
\affiliation{School of Physics and Astronomy, University of Edinburgh, Peter Guthrie Tait Road, Edinburgh, EH9 3FD, UK}
\author{Matt Newton}
\affiliation{School of Biosciences, Faculty of Science, University of Sheffield, Sheffield, S10 2TN, UK}
\author{Erin Cutts}
\affiliation{School of Biosciences, Faculty of Science, University of Sheffield, Sheffield, S10 2TN, UK}
\author{Davide Michieletto}
\thanks{corresponding author, davide.michieletto@ed.ac.uk}
\affiliation{School of Physics and Astronomy, University of Edinburgh, Peter Guthrie Tait Road, Edinburgh, EH9 3FD, UK}
\affiliation{MRC Human Genetics Unit, Institute of Genetics and Cancer, University of Edinburgh, Edinburgh EH4 2XU, UK}
\author{Antonio Valdés}
\thanks{corresponding author, antonio.valdes-gutierrez@uni-wuerzburg.de}
\affiliation{Biochemistry and Cell Biology, Biocenter, Julius-Maximilians-Universität of Würzburg, Wurzburg, Germany}

\begin{abstract}
\textbf{Type II Topoisomerases (topo II) are critical to simplify genome topology during transcription and replication. They identify topological problems and resolve them by passing a double-stranded DNA segment through a transient break in another segment. The precise mechanisms underpinning topo IIs ability to maintain a topologically simple genome are not fully understood. Here, we investigate how binding kinetics affects the resolution of two distinct forms of topological entanglement: decatenation and torsional relaxation. First, by single-molecule measurements, we quantify how monovalent cation concentration affects the dissociation rate of topo II from DNA. Second, we discover that increasing dissociation rates accelerate decatenation while slowing down relaxation catalytic activities. Finally, by using molecular dynamics simulations, we uncover that this opposite behaviour is due to a trade-off between search of target through facilitated diffusion and processivity of the enzyme in catenated versus supercoiled DNA. Thus, our findings reveal that a modulation of topo II binding kinetics can oppositely regulate its topological simplification activity, and in turn can have a significant impact \textit{in vivo}.
}
\end{abstract}

\maketitle

Type II topoisomerases are a family of proteins involved in the topological regulation of genomes across life forms~\cite{Wang2002}. Type IIA topoisomerase (topo II) catalyses double-strand DNA passage through ATP-dependent process that involves temporary covalent bonds between its tyrosine residues and the 5’phosphate group in the backbone of a G-DNA segment, allowing a T-DNA segment to pass through~\cite{ROCA1992, Bates2005,Bates2011}. 
The crucial role played by topo II is both well established and fascinating, as its ability to manage DNA entanglements appears essential for complex life~\cite{Wang2002}. However, the precise biophysical principles governing topo II–mediated topological simplification, especially in dense and crowded conditions of the nucleus or nucleoid, are poorly understood~\cite{Pommier2022,McKie2021}. 
While topo II can reduce the topological complexity of naked DNA below equilibrium in dilute conditions \emph{in vitro}~\cite{Rybenkov1997, Hardin2011, Dalvie2022}, it can also create catenated or knotted structures in dense conditions~\cite{Kim2013,Krajina2018} or in the presence of polycations~\cite{Krasnow1982}. Given the dense and entangled state of chromatin within the nucleus, how topo II maintains genomic simplicity \textit{in vivo} remains unclear.
Experiments on yeast minichromosomes~\cite{Valdes2018} and human chromatin~\cite{Goundaroulis2020,Siebert2017} suggest that genomic topological complexity \textit{in vivo} is lower than expected at equilibrium. 
Among the potential mechanisms to rationalise these observations are that topo IIs work in synergy with partner proteins, such as SMCs~\cite{Orlandini2019pnas,Dyson2020, McKie2021}, or that they are post-translationally modified in order to change their binding and phase separation properties~\cite{Dang1994,Jeong2022}. The latter mechanism suggests a potential mode of regulation of topo II catalytic activity based on its binding kinetics~\cite{Jeong2022,Michieletto2022nar}. 

\begin{figure}[h!]
	\centering	\includegraphics[width=0.45\textwidth]{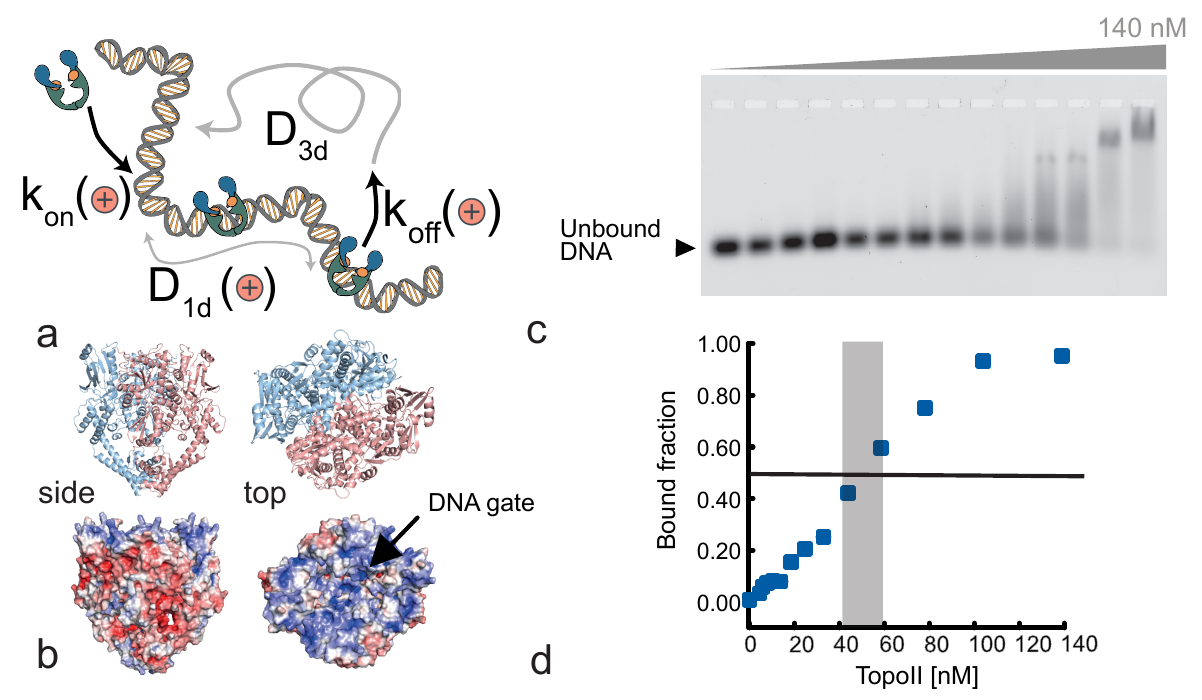}
	\caption{\textbf{a.} Topo II search on DNA, including 1D and 3D diffusion and intersegmental jumps. On and off rates, and 1D diffusion constant are expected to be salt dependent. \textbf{b.} Topo II present exposed positive surface charges that stabilise the binding with negatively charged DNA substrate. Its intrinsically disordered CTD also display positive charges that can bind to dsDNA (see SI). red = negative, blue = positive charges. \textbf{c.} Electrophoretic mobility shift assay (EMSA) performed on a 51 bp dsDNA varying topo II concentration from 0 to 140nM, at 125 mM NaCl. \textbf{d.} Quantification of the bound dsDNA fraction in the EMSA, returning an approximate $k_D \simeq 50$ nM. }
	\label{fig:Topo2_CTD}
\end{figure}

\begin{figure}[t!]
	\centering
	\includegraphics[width=0.45\textwidth]{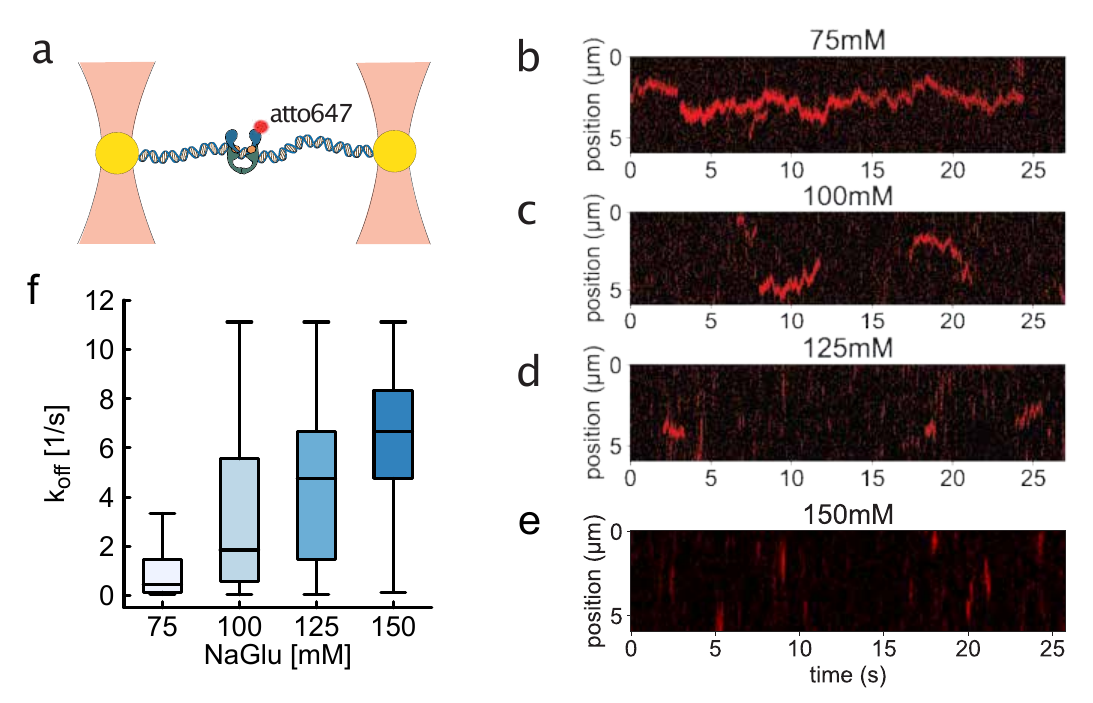}
	\caption{\textbf{Binding kinetics of topo II is modulated by monovalent cations.} \textbf{a} Diagram of force-stretched DNA. DNA tethered between trapped beads, topo II complex is labeled using atto647. \textbf{c-f} Kymographs of atto647-labelled topo II binding, diffusing and unbinding from a stretched $\lambda$DNA using optical tweezers at 75-150 mM [NaGlu]. \textbf{g} Quantification of inverse dwell time, or \koff, as a function of monovalent cation concentration. Each concentration includes analysis of more than 100 single-molecule tracks. All pair-wise distributions are significantly different (p value $< 0.01$).}
	\label{fig:restime}
\end{figure}

Binding kinetics is important for target search on coiled substrates like DNA~\cite{Lomholt2009}, where proteins can perform diffusion in 1D, 3D and can perform intersegmental jumps (Fig.~\ref{fig:Topo2_CTD}a). Faster binding and unbinding kinetics of topo II has been computationally suggested to yield faster topological simplification rates on knotted DNA substrates due to an effective ``enhanced 3D sampling'' of essential crossings~\cite{Michieletto2022nar}. However, it remains to be experimentally determined whether all catalytic activities of topo IIs are favoured by faster (un)binding kinetics. Motivated by this open question, in this work we tested topo II topological relaxation and decatenation activities on supercoiled and catenated DNA substrates to determine how they are affected by its (un)binding kinetics.

First, we determined that topo II binding kinetics can be modulated by varying monovalent cations within a physiological range. By screening negative charges along the DNA, monovalent cations reduced the binding affinity of topo II to the DNA and in turn accelerated its unbinding rate. We thus leveraged this modulation and performed topological simplification assays on catenated and supercoiled structures in the presence of different amounts of monovalent cations. Surprisingly, we observed opposite effects: decatenation was accelerated by higher cation concentration, while relaxation was slowed down. We rationalised these observations by performing Molecular Dynamics simulations of topologically constrained or catenated DNA plasmids under the action of a dynamically binding topo II, mimicking alternating 1D and 3D diffusion. Our simulated results – in line with the experimental findings – suggest that whilst efficient decatenation benefits from fast binding kinetics, torsional relaxation is favoured by more processive topo II with longer binding times.

\section{Results}

\begin{figure*}[t!]
	\centering
	\includegraphics[width=0.95\textwidth]{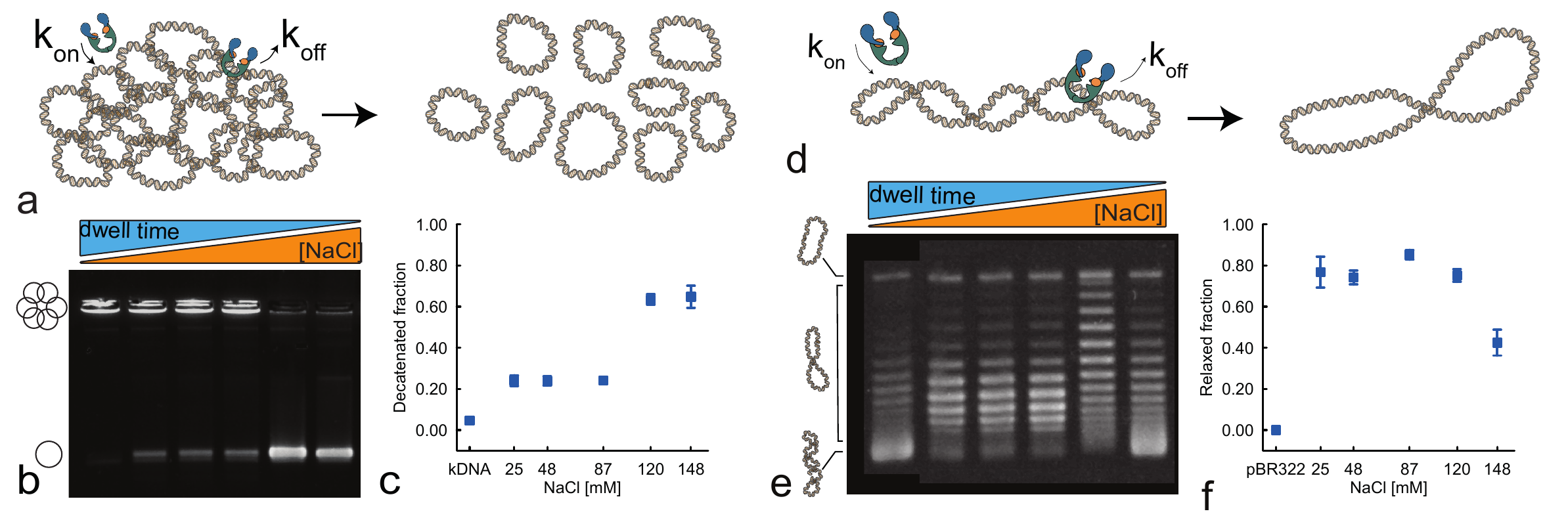}
	\caption{\textbf{Topoisomerase decatenation and relaxation activities are oppositely regulated by salt concentration.} \textbf{a.} To quantify decatenation activity we monitor the disassembly of kDNA as a function of various [NaCl]. \textbf{b.} Gel electrophoresis of 20 ng/ul kDNA after 30 minutes incubation with 3.45 ng/ul of topo II. The first lane shows the control sample without protein, and the following five lanes correspond to reactions performed in buffers containing [NaCl] = 25, 48, 87, 120, and 148 mM, respectively. \textbf{c.} Quantification of decatenated minicircles as a function of [NaCl]. \textbf{d.} To quantify the relaxation activity we monitor the topology of a negatively supercoiled pBR322 plasmid as a function of various [NaCl]. \textbf{e.} Chloroquine gel electrophoresis of 20 ng/ul pBR322 after 30 minutes incubation with 3.45 ng/ul of topo II. As in panel b, the first lane corresponds to the control sample without protein, and the remaining lanes show reactions carried out at [NaCl] = 25, 48, 87, 120, and 148 mM. Notably the amounts of supercoiled and relaxed DNA at 120mM and 87mM NaCl are comparable, nevertheless the supercoil distributions are visibly different, showing a slowed relaxation effect at higher NaCl concentrations. \textbf{f.} Quantification of relaxed topoisomer fraction as a function of [NaCl].}
	\label{fig:exp}
\end{figure*}

\subsection{Topo II binding kinetics is modulated by monovalent cations}

Monovalent cations change the binding affinity of DNA-binding proteins by screening electrostatic interactions. More specifically, they can significantly increase the dissociation rate (\koff) of proteins from DNA whilst remaining within physiologically relevant concentrations (e.g., 50-150 mM NaCl)~\cite{Kamar2017,Jones2016}. Topo IIs display three principal positively charged surfaces that are thought to interact with, and stabilise, the nucleo-protein complex. A major patch lies along the DNA cleavage-rejoining domain, or DNA gate, which binds the G-DNA segment (Fig.~\ref{fig:Topo2_CTD}b) and two additional patches are located on the intrinsically disordered C-terminal domains (CTD) (Fig.~S1a). We thus decided to quantify the (un)binding kinetics of yeast topo II on DNA for different physiological levels of monovalent cations (Na$^+$). 
First, we precisely measured the dissociation kinetics \koff at single molecule resolution. We used confocal microscopy to visualise single fluorescently labelled topo II binding, diffusing and unbinding on $\lambda$DNA stretched using optical tweezers (see SI for details, Fig.~\ref{fig:restime}a). We performed the same experiments at different levels of [Na$^+$] and we then analysed the kymographs (over 100 tracks for each condition), extracted the length of the single traces, and finally determined the dwell time, or inverse dissociation constant \koff (Fig.~\ref{fig:restime}b-e). The values of \koff measured range within $0.1-10 \;s^{-1}$ and significantly increase with higher concentrations of monovalent cations, in line with what observed for other DNA-binding proteins \cite{Vanderlinden2022, Charvin2005a}. 
We also measured the apparent dissociation constant $k_D$ via electrophoretic mobility shift assay (EMSA) on a short (51 bp) dsDNA (Fig.~\ref{fig:Topo2_CTD}c,d).
We can compare these measurements with estimates from a Smoluchowski collision equation; the on-rate can be estimated as $k_{on} \simeq 4 \pi (D_1 + D_2) (a_1 + a_2) \simeq 5\times 10^9 M^{-1}  s^{-1}$ with $D_1 = k_BT/(3 \pi \eta_w a_1)\simeq 10 \mu m^2/s$, $D_2 \ll D_1$, $a_1=40$ nm the size of topo II~\cite{Hizume2007,Schultz1996}, $a_2=10$ nm the size of the target DNA site and where we used water viscosity $\eta_w = 1 $ mPa s. This yields a dissociation constant $k_{\textrm off} = k_D k_{on} = 50 s^{-1}$ which is within the same order of magnitude of the ones measured in single molecule experiments (Fig.~\ref{fig:restime}f).

\subsection{Faster binding kinetics accelerates decatenation and slows down relaxation}

Having quantified how the dwell time $\tau_{off} = 1/k_{off}$ changes with the concentration of monovalent cations in solution, we then asked how this parameter affected the catalytic activity of topo II to perform topological decatenation of catenated, kinetoplast DNA (kDNA)~\cite{Chen1995,He2023} (Fig.~\ref{fig:exp}a). To this end, we incubated \textit{Crithidia fasciculata} kDNA, a structure made of thousands of catenated DNA rings \cite{Chen1995,He2023}, with yeast topo II at various concentrations of monovalent cations. The enzyme concentration was set to be comparable to the substrate molarity, and reaction conditions were selected to ensure that decatenation did not reach full topological simplification; reactions were then stopped and the samples were run on a gel to separate catenated from single-circle topologies. We observe that the decatenation rate is sped up by higher levels of NaCl, corresponding to shorter dwell times of topo II on DNA (Fig.~\ref{fig:exp}b-c). This result is consistent with previous numerical results, where different modes of topo II (un)binding to DNA substrates were found to enhance topo II unknotting activity~\cite{Michieletto2022nar}. 

We then performed topological relaxation experiments where a 4.5 kb negatively supercoiled plasmid (pRB322) was incubated with yeast topo II and varying [NaCl] (Fig.~\ref{fig:exp}d). As for the decatenation assay, the enzyme concentration was kept comparable to the substrate molarity, and conditions were chosen to prevent the reaction from reaching full topological relaxation. We then ran the samples through a gel in presence of 0.6 $\mu$g/ml of chloroquine and quantified the distribution of topoisomers (Fig.~\ref{fig:exp}e-f). The enzyme concentration was set to be comparable to the substrate molarity, while reaction conditions were selected to ensure that, given the enzyme’s activity, neither decatenation nor relaxation proceeded to full topological simplification. This allowed us to resolve intermediate topological states and evaluate the salt dependence of activity. Unexpectedly, we found that topo II was less efficient at relaxing supercoiling at larger salt concentrations. In fact, while at low [NaCl] the topoisomer distributions were tightly peaked around the relaxed $Lk_0$, at larger [NaCl] we could observe a very broad distribution (5th lane in Fig.~\ref{fig:exp}b). This observation is consistent with the short dwell time of topo II on DNA and suggests that the enzyme performs short rounds of catalytic activity in these conditions. On the other hand, it is conceptually opposite to what we observed in the decatenation experiment. These results were consistent in both sodium chloride and sodium glutamate and did not depend on the presence of the CTD (Fig. S6), even though the CTD is a strong DNA-binding site (Fig. S1b).

\subsection{MD simulations explain the opposite catalytic regulation by monovalent cations} 

\begin{figure*}[t!]
	\includegraphics[width=0.95\textwidth]{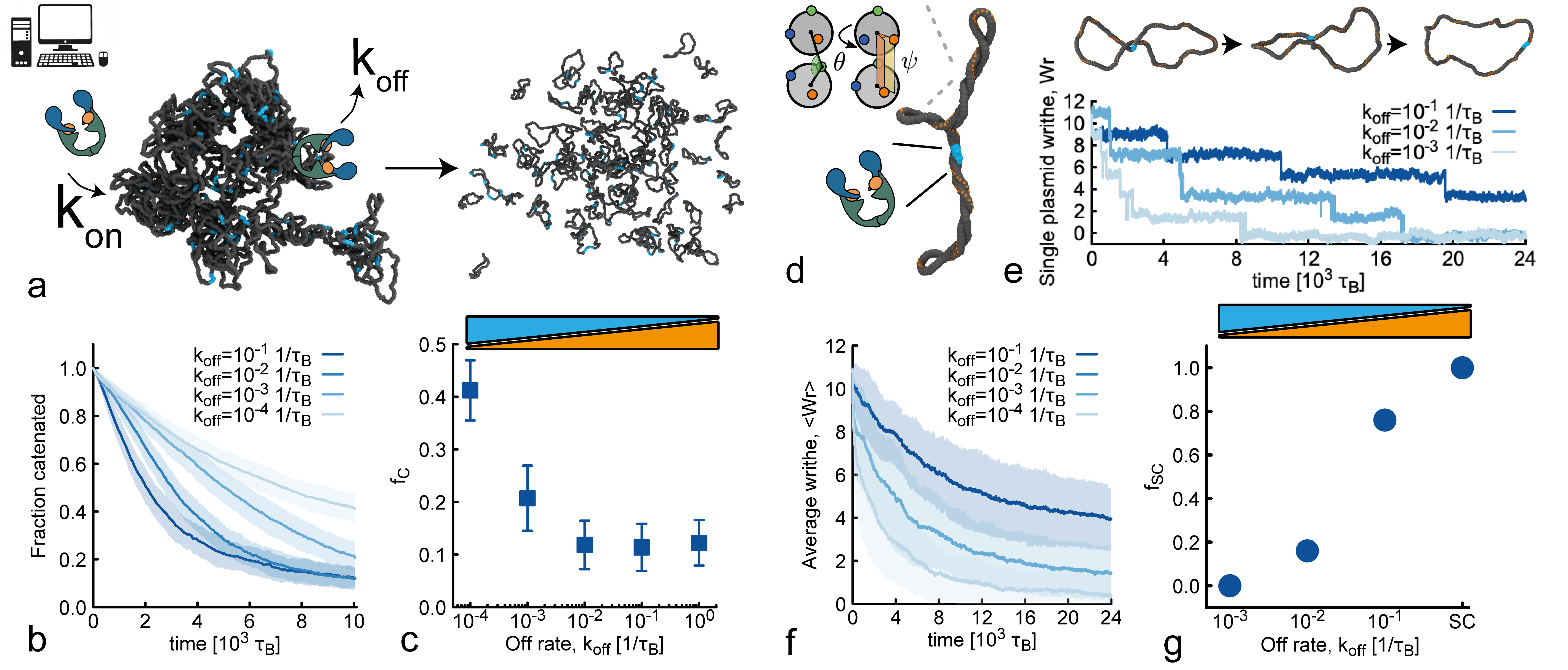}
	\caption{\textbf{Molecular Dynamics simulations with salt-dependent \koff capture the opposite topological regulation.} \textbf{a.} Snapshots from MD simulations of a model kDNA from Ref.~\cite{He2023} before and after decatenation reaction. \textbf{b.} Fraction of catenated minicircles as a function of time and for different values of off-rate. Shaded area represents the standard deviation over 50 independent replicas. \textbf{c.} Fraction of catenated minicircles at fixed time ($10^4 \tau_B$) and as a function of \koff showing enhanced decatenation at larger off-rate. \textbf{d.} Snapshot from MD simulations of a supercoiled DNA chain (model from Ref.~\cite{Brackley2014supercoil,Smrek2021}. \textbf{e.} Step-wise change in writhe monitored through the simulation. Snapshots at the top, plot of writhe at the bottom. \textbf{f.} Average writhe $\langle \mathcal{W} \rangle$ as a function of time and for different values of off-rate. The shaded area represents the standard deviation across 50 independent replicas. \textbf{g.} Fraction of supercoiled topoisomers (defined as those with $\mathcal{W}/N > 0.05$) as a function of off-rate (SC = fully supercoiled state). }
	\label{fig:sim}
\end{figure*}

To rationalise our experiments, we decided to perform Molecular Dynamics (MD) simulations of a coarse-grained bead-spring polymer model of DNA under the action of kinetically binding topo II. 

First, we simulated a 1 $\mu$m$^2$ patch of kinetoplast DNA made of 90 linked DNA minicircles, each made of 2.5 kb or 85 beads, whose topology was inferred from AFM data (Ref.~\cite{He2023}). We then run a Langevin simulation (implicit solvent, see SI) including a sub-stoichiometric number of topo IIs ($N_t=80$), which were modelled by turning a 5-beads segment of the polymer into a ``phantom'' segment. During the simulation, the polymer chains undergo Brownian motion and most of the beads interact via a uncrossable, purely repuslive potential. The $N_t=80$ ``phantom'' segments representing topo II are the only segments that can be freely crossed and are transiently added/removed from the polymers stochastically, with typical on- and off-rates that mimic different [NaCl] conditions (Fig.~\ref{fig:sim}a). More specifically, we fixed the on-rate to 0.001 $1/\tau_B$ and increased the off-rate from $10^{-4} 1/\tau_B$ to $1/\tau_B$ to capture shorter dwell times at increasing salt concentrations.  Additionally, we assume that topo II does not perform 1D diffusion along the rings. During the simulation we track the topology of the network via the pairwise linking number 
\begin{equation}
   \mathcal{L}_{ij} = \oint_{\gamma_i} \oint_{\gamma_j} ds \, ds^\prime \dfrac{(\bm{t}(s) \times \bm{t}(s^\prime)) \cdot (\bm{r}(s) - \bm{r}(s^\prime))}{|\bm{r}(s) - \bm{r}(s^\prime)|^3} \, ,
    \label{eq:wr}  
\end{equation}
where $\gamma_1$ and $\gamma_2$ are simulated polymer rings, $\bm{t}(s)$ and $\bm{r}(s)$ are the tangent and position of the polymers at arclength $s$. Given $\mathcal{L}_{ij}$ for all the pairs of rings, we can calculate how many rings belong to the largest interlocked component in the system. Because we start from a fully catenated structure, the fraction of catenated rings $f_C$ will start from one and decay in time. By averaging over 50 independent replicas of the system we thus obtain the average fraction of catenated $\langle f_C(t)\rangle$ plotted in Fig.~\ref{fig:sim}b. These curves can be used to directly compare our simulations with the gels in Fig.~\ref{fig:exp}b-c. Specifically, we compare $\langle f_C\rangle$ at fixed runtime $\bar{t}$ and for different values of \koff and observe that the larger the off-rate the larger the smaller the catenated fraction -- and hence the larger the decatenated fraction -- as seen in experiments.   
Interestingly, and in line with experiments, we also observe that $\langle f_C(\bar{t})\rangle$ plateaus at large \koff suggesting that the decatenation process is reaction limited, i.e. a minimum topo II dwell time is required to perform successful decatenation. In other words, it is not possible to accelerate the decatenation process indefinitely after some value of the off-rate, which is required to catalyse at least one double strand passage.

Having observed that our simulations can recapitulate the experimental results for the decatenation activity of topo II we decided to computationally test its relaxation activity. To model supercoiled DNA we employ a coarse-grained ``twistable'' polymer (see Refs.~\cite{Brackley2014supercoil,smrek2021topological} for details). 
We implement a dynamically binding/unbinding topo II by transiently creating/deleting a ``phantom'' segment from the chain (see Fig.~\ref{fig:sim}a and Methods). We maintain the on-rate constant and increase \koff to capture the faster unbinding at larger [Na$^+$].  The topological relaxation of our supercoiled polymer can be quantified by measuring its writhe~\cite{Klenin1991,Michieletto2016tree}
\begin{equation}
   \mathcal{W} = \oint_\gamma \oint_{\gamma} ds \, ds^\prime \dfrac{(\bm{t}(s) \times \bm{t}(s^\prime)) \cdot (\bm{r}(s) - \bm{r}(s^\prime))}{|\bm{r}(s) - \bm{r}(s^\prime)|^3} \, ,
    \label{eq:wr}  
\end{equation}
where $\bm{t}(s)$ and $\bm{r}(s)$ are the tangent and position of the curve $\gamma$ at arclength $s$. The relaxation proceeds by allowing the polymer backbone to cross through itself, as shown in Fig.~\ref{fig:sim}b. Each strand-crossing event yields a step-wise change in writhe $\Delta \mathcal{W}=-2$. Note that the writhe is not an integer and can fluctuate as long as the sum of twist and writhe of single chains remain constant (CFW theorem~\cite{Tubiana2024}). By averaging over 50 independent replicas we then obtain relaxation curves $\langle \mathcal{W} (t) \rangle$ plotted in Fig.~\ref{fig:sim}f. Surprisingly, and opposite to what we saw for the decatenation reaction, we observe that larger \koff leads to slower relaxation (Fig.~\ref{fig:sim}g). However, this slowing down is consistent with the experiments in Fig.~\ref{fig:exp} and demonstrates opposite regulation of catalytic activities of topo II simply by a change in monovalent cation concentration.  

\section{Discussion}

In this work we observed a significant and opposite regulation of topo II topological decatenation and relaxation activities as a function of monovalent cation concentration.  

We rationalise the observed behaviour as follows: while at large [Na$^+$] topo II spends more time performing unbound 3D search, at small [Na$^+$] topo II spends more time bound to the DNA. Thus, in the regime of low monovalent cation concentration, topo II is highly processive and performs several rounds of strand-passage without leaving the substrate. In contrast, at high monovalent cation concentration topo II is less processive, quickly dissociating from DNA and spending more time performing facilitated 3D diffusion. 

These different binding kinetics have profound effects on topo II catalytic activities. Supercoiled plasmids have highly localised entanglements and are therefore rapidly relaxed by a highly processive enzyme that remains bound long enough to perform sequential strand-passage events.
On the contrary, entanglements resulting from catenation are typically delocalised across the substrate and therefore too long binding times yield inefficient decatenation because after a strand passage, no further entanglements are present at a given location. Thus, efficient decatenation is achieved through rapid binding–unbinding dynamics that allow topo II to disengage quickly and relocate to remaining entanglements. Our results thus reveal an intrinsic kinetic trade-off: torsional relaxation is favoured by prolonged binding, whereas decatenation is accelerated by faster turnover.

We argue that this opposite regulation may have physiological relevance. Indeed, tuning the binding affinity of topo II to DNA, for example by post-translational modifications~\cite{Jeong2022}, it is possible to oppositely regulate topo II decatenation and relaxation activities.

This finding is consistent with the view that topoisomerases are recruited and controlled through protein–protein interactions and post-translational modifications that localize them to specific genomic regions or cell-cycle stages. For example, recent studies show that the oncoprotein MYC can form a complex with human topo II$\alpha$, enhancing topoisomerase retention at regions experiencing elevated transcription-induced supercoiling~\cite{Das2022,Cameron2025}. In this framework, increasing enzyme dwell time at highly supercoiled loci would promote efficient relaxation, consistent with our observations under low-salt conditions. Conversely, during stages of the cell cycle that demand rapid decatenation, such as metaphase, when sister chromatids must be fully separated before anaphase, excessively long enzyme dwell times could be detrimental, increasing the likelihood of unwanted re-catenation events. Our results under high-salt conditions, which shorten DNA residence time and accelerate decatenation, suggest a biophysical basis for how cells might avoid such outcomes. Together, our findings provide mechanistic insight into how topo II activity can be tuned through its binding kinetics, offering a unified explanation for how cells could differentially regulate supercoil relaxation and chromosome decatenation. By demonstrating how subtle shifts in dwell time produce qualitatively distinct topological outcomes, this work highlights binding kinetics as a central and underappreciated layer of topo II regulation \textit{in vivo}.




\section{Materials and Methods}

\subsection{Protein expression and purification} 
Full-length and C-terminal domain truncated topo II were expressed in yeast (see SI for detailed expression and purification protocol). Ybbr peptide tag on full-length topo II was labelled with Sfp phospho-pantetheinyl transferase (P9302, New England Biolabs) and CoA-ATTO647N (see SI for detailed fluorescent labelling protocol).

\subsection{Electrophoretic mobility assay} 
Electrophoretic mobility assay (EMSA) was carried out by mixing a 6-FAM labelled 51 bp dsDNA with protein at the concetrations indicated in the figure in 40 mM TRIS-HCl pH 7.5, 125 mM NaCl, 5 mM MgCl2, 10\% glycerol, 1 mM DTT. After 10 min incubation at room temperature ($\sim$ 25$^\circ$ C), free DNA and DNA-protein complexes were resolved by electrophoresis for 1 h at 4 V/cm on 0.75\% (w/v) TAE-agarose gels at 4$^\circ$C. We then imaged the 6-FAM labelled DNA using a gel scanner (see SI for more details).

\subsection{DNA relaxation assay} 

200 ng of pBR322 (NEB) plasmid were incubated with 34.5 ng of topoII-$\Delta$CTD in a 10 ul volume containing 50mM Tris-HCl (pH 7.5), [25,48,87,120,148] mM NaCl, 5 mM MgCl2, 1 mM DTT. ATP was added to 1mM and the mixture was incubated at 37$^\circ$C for 30 min. Reactions were stopped with the addition of 20mM EDTA, 1\% (w/v) sodium dodecyl sulphate (SDS). Subsequently, 0.8 U of proteinase K were added, followed by incubation for 1 h at 65$^\circ$C. Reaction samples were loaded onto 1\% (w/v) agarose gels. DNA electrophoresis was performed at 2 V/cm for 15 h in 1X TAE buffer (40 mM Tris, 20 mM acetate, 1 mM EDTA) containing 0.6 $\mu$g/mL chloroquine. Gels were then stained with SYBR Gold (Thermo Fisher Scientific) and visualized using a Syngene transilluminator. Gel images were analyzed using Fiji~\cite{Fijipaper} (see SI).

\subsection{DNA decatenation assay} 
The protocol used for the decatenation assay is the same as the one used for the relaxation assay, except for the following differences: Instead of pBR322 plasmid, 200 ng of Kinetoplast DNA (Inspiralis) are added to the reactions. No chloroquine is used for the electrophoresis.

\subsection{Molecular Dynamics simulations}
Molecular Dynamics simulations were performed in LAMMPS~\cite{Plimpton1995} in NVT ensemble using a Langevin thermal bath and custom fixes that changed the type of beads during the course of the simulations. In both relaxation and decatenation simulations we considered sub-stoichiometric number of topo IIs, in line with experiments. Topo IIs were modelled by turning a 5-beads segment of the polymer into a ``phantom'' segment. All other beads along the polymers interacted via a uncrossable, purely repulsive potential. The ``phantom'' segments were transiently added/removed from the polymers stochastically, with typical on- and off-rates that mimic different [NaCl] conditions (Fig.~\ref{fig:sim}a). More specifically, we fixed the on-rate to 0.001 $1/\tau_B$ and increased the off-rate from $10^{-4} 1/\tau_B$ to $1/\tau_B$ to capture shorter dwell times at increasing salt concentrations. See SI for more details of the simulations.

\subsection{Single molecule assay}

Single molecule assay was conducted using the Lumicks C-Trap instrument BA105, equipped with integrated confocal microscopy and microfluidics (Fig.~\ref{fig:restime}a, SI Fig.2a). The fluidics was cleaned by flowing ~500 ul 2\% Hellmanex over 40 minutes followed by flowing 2 x ~0.5 ml MQ H2O and finally 0.5 ml Optical tweezers Buffer, OTB (50mM Tris–HCl (pH 7.5), [75,100,150,200] mM NaGlu, 5 mM MgCl2, 1 mM DTT). Protein channels (Channels 4 and 5) were passivated with 0.5 ml Casein (0.01\% w/v) and BSA (0.005\% w/v) in OTB flowed through over a 30 minute period. $\lambda$-DNA was prepared by ligation of biotinylated caps (as previously described \cite{NEWTON20233533, King2019}. Fluidics channels were prepared as follows: Channel 1, 0.005\% w/v  4.38 um SPHERO™Streptavidin coated polystyrene beads (SVP-40-5) in OTB; Channel 2, biotinylated DNA diluted 200-fold in OTB; Channel 3, OTB; Channels 4 and 5, topo2-atto647 at 2-3 nM in OTB. OTB in Channels 4 and 5 was prepared with varying NaGlu concentrations (75, 100, 150, or 200 mM) while kept at constant NaGlu concentration of 100 mM in the other channels. DNA was and tethered between optically trapped beads using the laminar flow cell. Prior to each experiment, a force-extension curve was measured from 0–55 pN to confirm the presence of a single intact DNA molecule. DNA was held at 5 pN before moving into the channel containing protein. Confocal imaging was performed using 5\% excitation power at 647 nm, with emission detected through a 680/42 nm band-pass filter; images were acquired with a pixel size of 100 nm and a pixel dwell time of 0.1 ms, using a line width of 25.5 $\mu$m and a line time of 30 ms (See SI for data analysis).

\section{Data availability}
All the information are included in the Supplementary Information. Codes are deposited at \url{https://git.ecdf.ed.ac.uk/taplab}.

\section{Acknowledgements}
D.M. acknowledges the Royal Society and the European Research Council (grant agreement No 947918, TAP) for funding. We thank Nesibe Durmaz for technical assistance. The authors also acknowledge the contribution of the COST Action Eutopia, CA17139. For the purpose of open access, the author has applied a Creative Commons Attribution (CC BY) licence to any Author Accepted Manuscript version arising from this submission. M.D.N. is supported by a Wellcome Early Career Award (225139/Z/22/Z). The LUMICKS C-trap instrument was funded through MRC Equipment Grant MC\_PC\_APP28599 and The Henry Royce Institute for Advanced Materials, funded through EPSRC grants EP/R00661X/1, EP/S019367/1, EP/P02470X/1 and EP/P025285/1. The work was funded by Royce Access Scheme SHEF-YR10-003\_SREAS25/006. We would like to acknowledge Xinyue Chen for technical support at Royce@Sheffield.

\bibliography{biblio_fin,t2bib}

\end{document}